\documentclass{ian}

\usepackage{dsfont}
\usepackage{iantheo}
\usepackage{largearray}
\usepackage{point}
\usepackage{etoolbox}

\begin{document}

\pagestyle{empty}

\hbox{}
\hfil{\bf\LARGE
The Simplified approach to the Bose gas\par
\medskip
\hfil without translation invariance
}
\vfill

\hfil{\bf\large Ian Jauslin}\par
\hfil{\it Department of Mathematics, Rutgers University}\par
\hfil{\tt\color{blue}\href{mailto:ian.jauslin@rutgers.edu}{ian.jauslin@rutgers.edu}}\par
\vskip20pt

\vfill

\hfil {\bf Abstract}\par
\smallskip

The Simplified approach to the Bose gas was introduced by Lieb in 1963 to study the ground state of systems of interacting Bosons.
In a series of recent papers, it has been shown that the Simplified approach exceeds earlier expectations, and gives asymptotically accurate predictions at both low and high density.
In the intermediate density regime, the qualitative predictions of the Simplified approach have also been found to agree very well with Quantum Monte Carlo computations.
Until now, the Simplified approach had only been formulated for translation invariant systems, thus excluding external potentials, and non-periodic boundary conditions.
In this paper, we extend the formulation of the Simplified approach to a wide class of systems without translation invariance.
This also allows us to study observables in translation invariant systems whose computation requires the symmetry to be broken.
Such an observable is the momentum distribution, which counts the number of particles in excited states of the Laplacian.
In this paper, we show how to compute the momentum distribution in the Simplified approach, and show that, for the Simple Equation, our prediction matches up with Bogolyubov's prediction at low densities, for momenta extending up to the inverse healing length.

\vfill

\tableofcontents

\eject

\setcounter{page}1
\pagestyle{plain}

\section{Introduction}
\indent
The Bose gas is one of the simplest models in quantum statistical mechanics, and yet it has a rich and complex phenomenology.
As such, it has garnered much attention from the mathematical physics community for over half a century.
It consists in infinitely many identical Bosons and is used to model a wide range of physical systems, from photons in black body radiation to gasses of helium atoms.
Whereas photons do not directly interact with each other, helium atoms do, and such an interaction makes studying such systems very challenging.
To account for interactions between Bosons, Bogolyubov\-~\cite{Bo47} introduced a widely used approximation scheme that accurately predicts many observables\-~\cite{LHY57} {\it in the low density} regime.
Even though Bogolyubov theory is not mathematically rigorous, it has allowed mathematical physicists to develop the necessary intuition to prove a wide variety of results about the Bose gas, such as the low density expansion of the ground state energy of the Bose gas in the thermodynamic limit\-~\cite{Dy57,LY98,YY09,FS20,BCS21,FS22}, as well as many other results in scaling limits other than the thermodynamic limit (see\-~\cite{Sc22} for a review, as well as, among many others, \cite{LSY00,LS02,NRS16,BBe18,BBe19,DSY19,BBe20,DS20,NT21,BSS22,BSS22b,HST22,NNe22}).
In this note, we will focus on the ground state in the thermodynamic limit.
\bigskip

\indent
In 1963, E.H.\-~Lieb\-~\cite{Li63,LS64,LL64} introduced a new approximation scheme to compute properties of the ground state of Bose gasses, called the {\it Simplified approach}, which has recently been found to yield surprisingly accurate results\-~\cite{CJL20,CJL21,CHe21,Ja22}.
Indeed, while Bogolyubov theory is accurate at low densities, the Simplified approach has been shown to yield asymptotically accurate results at both {\it low and high} densities\-~\cite{CJL20,CJL21} for interaction potentials that are of positive type, as well as reproduce the qualitative behavior of the Bose gas at intermediate densities\-~\cite{CHe21}.
In addition to providing a promising tool to study the Bose gas, the derivation of the Simplified approach is different enough from Bogolyubov theory that it may give novel insights into longstanding open problems about the Bose gas.
\bigskip

\indent
The original derivation of the Simplified approach\-~\cite{Li63} is quite general, and applies to any translation invariant system (it even works for Coulomb\-~\cite{LS64} and hard-core\-~\cite{CHe21} interactions).
In the present paper, we extend this derivation to systems that break translation invariance.
This allows us to formulate the Simplified approach for systems with external potentials, and with a large class of boundary conditions.
In addition, it allows us to compute observables in systems with translation invariance, but whose computation requires breaking the translation invariance.
We will discuss an example of such an observable: the momentum distribution.
\bigskip

\indent
The momentum distribution $\mathcal M(k)$ is the probability of finding a particle in the state $e^{ikx}$.
Bose gasses are widely expected to form a Bose-Einstein condensate, although this has still not been proven (at least for continuum interacting gasses in the thermodynamic limit).
From a mathematical point of view, Bose-Einstein condensation is defined as follows: if the Bose gas consists of $N$ particles, the average number of particles in the constant state (corresponding to $k=0$ in $e^{ikx}$) is of order $N$.
The {\it condensate fraction} is defined as the proportion of particles in the constant state.
The momentum distribution is an extension of the condensate fraction to a more general family of states.
In particular, computing $\mathcal M(k)$ for $k\neq 0$ amounts to counting particles that are {\it not} in the condensate.
This quantity has been used in the recent proof\-~\cite{FS20,FS22} of the energy asymptotics of the Bose gas at low density.
\bigskip

\indent
The main results in this paper fall into two categories.
First, we will derive the Simplified approach without assuming translation invariance, see Theorem\-~\ref{theo:simple}.
To do so, we will make the so-called ``factorization assumption'', on the marginals of the ground state wavefunction, see Assumption\-~\ref{assum:factorization}.
This allows us to derive a Simplified approach for a wide variety of situations in which translation symmetry breaking is violated, such as in the presence of external potentials.
Second, we compute a prediction for the momentum distribution using the Simplified approach.
The Simplified approach does not allow us to compute the ground state wavefunction directly, so to compute observables, such as the momentum distribution, we use the Hellmann-Feynman technique and add an operator to the Hamiltonian.
In the case of the momentum distribution, this extra operator is a projector onto $e^{ikx}$, which breaks the translation invariance of the system.
In Theorem\-~\ref{theo:Nk}, we show how to compute the momentum distribution in the Simplified approach using the general result of Theorem\-~\ref{theo:simple}.
In addition, we check that the prediction is credible, by comparing it to the prediction of Bogolyubov theory, and find that both approaches agree at low densities and small $k$, see Theorem\-~\ref{theo:Nk_bog}.
\bigskip

\indent
The rest of the paper is structured as follows.
In Section\-~\ref{sec:model}, we specify the model and state the main results precisely.
We then prove Theorem\-~\ref{theo:simple} in Section\-~\ref{sec:simple}, Theorem\-~\ref{theo:Nk} in Section\-~\ref{sec:Nk_proof}, and Theorem\-~\ref{theo:Nk_bog} in Section\-~\ref{sec:Nk_bog}.
The proofs are largely independent and can be read in any order.
\bigskip

\section{The model and main results}\label{sec:model}
\indent
Consider $N$ Bosons in a box of volume $V$ denoted by $\Omega_V:=[-V^{\frac13}/2,V^{\frac13}/2]^3$, interacting with each other via a pair potential $v\in L_{1}(\Omega_V^2)$ that is symmetric under exchanges of particles: $v(x,y)\equiv v(y,x)$.
The Hamiltonian acts on $L_{2,\mathrm{sym}}(\Omega_V^N)$ as
\begin{equation}
  \mathcal H:=
  -\frac12\sum_{i=1}^N\Delta_i
  +
  \sum_{1\leqslant i<j\leqslant N}v(x_i,x_j)
  +
  \sum_{i=1}^N P_i
  \label{ham}
\end{equation}
where $\Delta_i\equiv\partial_{x_i}^2$ is the Laplacian with respect to the position of the $i$-th particle and $P_i$ is an extra single-particle term of the following form: given a self-adjoint operator $\varpi$ on $L_2(\Omega_V)$,
\begin{equation}
  P_i:=\mathds 1^{\otimes i-1}\otimes \varpi\otimes\mathds 1^{\otimes N-i}
  .
  \label{Ppi}
\end{equation}
For instance, if we take $\varpi$ to be a multiplication operator by a function $v_0$, then $\sum_i P_i$ is the contribution of the external potential $v_0$.
Or $\varpi$ could be a projector onto $e^{ikx}$, which is what we will do below to compute the momentum distribution.
Because $P_i$ acts on a single particle, it breaks translational symmetry as soon as it is not constant.
\bigskip

\indent
We may impose any boundary condition on the box, as long as the Laplacian is self-adjoint.
We will consider the thermodynamic limit, in which $N,V\to\infty$, such that
\begin{equation}
  \frac NV=\rho
\end{equation}
is fixed.
We consider the ground state $\psi_0$, which is the eigenfunction of $\mathcal H$ with the lowest eigenvalue $E_0$:
\begin{equation}
  \mathcal H\psi_0=E_0\psi_0
  .
  \label{eigval}
\end{equation}
(It is a standard argument to prove that $\psi_0$ exists, and is both real and non-negative.)
\bigskip

\indent
In order to take the thermodynamic limit, we will assume that $v$ is uniformly integrable in $V$:
\begin{equation}
  |v(x,y)|\leqslant \bar v(x,y)
  ,\quad
  \int_{\mathbb R^3} dy\ \bar v(x,y)\leqslant c
  \label{intv}
\end{equation}
where $\bar v$ and $c$ are independent of $V$.
In addition, we assume that, for any $f$ that is uniformly integrable in $V$,
\begin{equation}
  \int dx\ \varpi f(x)\leqslant c
  .
  \label{bound_varpi}
\end{equation}
\bigskip

\subsection{The Simplified approach without translation invariance}\label{sec:general}
\indent
The crucial idea of Lieb's construction\-~\cite{Li63} is to consider the wave function $\psi$ as a probability distribution, instead of the usual $|\psi|^2$.
Since $\psi\geqslant 0$, this can be done by normalizing $\psi$ by its $L_1$ norm.
We then define the $i$-th marginal of $\psi$ as
\begin{equation}
  g_i(x_1,\cdots,x_i)
  :=
  \frac{\int\frac{dx_{i+1}}V\cdots\frac{dx_N}V\ \psi(x_1,\cdots,x_N)}{\int\frac{dy_{1}}V\cdots\frac{dy_N}V\ \psi(y_1,\cdots,y_N)}
  \equiv
  V^i\frac{\int dx_{i+1}\cdots dx_N\ \psi(x_1,\cdots,x_N)}{\int dy_{1}\cdots dy_N\ \psi(y_1,\cdots,y_N)}
  .
  \label{gdef}
\end{equation}
In particular, for $i\in\{2,\cdots,N\}$,
\begin{equation}
  \int\frac{dx_i}V\ g_i(x_1,\cdots,x_i)=g_{i-1}(x_1,\cdots,x_{i-1})
  ,\quad
  \int\frac{dx}V\ g_1(x)=1
  .
  \label{grec}
\end{equation}
Because of the symmetry of $\psi$ under exchanges of particles, $g_i$ is symmetric under $x_i\leftrightarrow x_j$.
\bigskip

\indent
Inspired by\-~\cite{Li63}, we will make the following approximation.
\bigskip

\theoname{Assumption}{Factorization}\label{assum:factorization}
  For $i=2,3,4$,
  \begin{equation}
    g_i(x_1,\cdots,x_i)
    =
    \prod_{1\leqslant j<l\leqslant i}
    W_i(x_j,x_l)
    \label{g_factorized}
  \end{equation}
  with
  \begin{equation}
    W_i(x,y)=f_i(x)f_i(y)(1-u_i(x,y))
    \label{W_fact}
  \end{equation}
  in which $f_i$ and $u_i$ are bounded independently of $V$ and $u_i$ is uniformly integrable in $V$:
  \begin{equation}
    |u_i(x,y)|\leqslant\bar u_i(x,y)
    ,\quad
    \int dy\ \bar u_i(x,y)\leqslant c_i
    \label{assum_bound}
  \end{equation}
  with $c_i$ independent of $V$.
  We further assume that, for $i=1,2,3$,
  \begin{equation}
    \lim_{V\to\infty}\int dx_i\ \Delta_{x_i} g_i(x_1,\cdots,x_i)=0
  \end{equation}
  in other words, these boundary terms vanish in the thermodynamic limit.
\endtheo
\bigskip

In other words, $g_i$ factorizes exactly as a product of pair terms $W_i$.
The $f_i$ in $W_i$ allow for $W_i$ to be modulated by a slowly varying density, which is the main novelty of this paper compared to\-~\cite{Li63}.
The inequality\-~(\ref{assum_bound}) ensures that $u_i$ decays sufficiently fast on the microscopic scale.
Note that, by the symmetry under exchanges of particles, $u_i(x,y)\equiv u_i(y,x)$.
\bigskip

\indent
Here, we use the term ``assumption'' because it leads to the Simplified approach.
However, it is really an {\it approximation} rather than an assumption: this factorization will certainly not hold true exactly.
At best, one might expect that the assumption holds approximately in the limit of small and large $\rho$, and for distant points, as numerical evidence suggests in the translation invariant case.
In the present paper, we will not attempt a proof that this approximation is accurate, and instead explore its consequences.
Suffice it to say that this approximation is one of {\it statistical independence} that is reminiscent of phenomena arising in statistical mechanics when the density is low, that is, when the interparticle distances are large.
In the current state of the art, we do not have much in the way of an explanation for why this statistical independence should hold; instead, we have extensive evidence, both numerical\-~\cite{CHe21} and analytical\-~\cite{CJL20,CJL21}, that this approximation leads to very accurate predictions.
\bigskip

\indent
The equations of the Simplified approach are derived from Assumption\-~\ref{assum:factorization}, using the eigenvalue equation\-~(\ref{eigval}) along with
\begin{equation}
  \int\frac{dx}V\ g_1(x)=1
  \label{g11}
\end{equation}
\begin{equation}
  \int\frac{dy}V\ g_2(x,y)=g_1(x)
  \label{g2g1}
\end{equation}
\begin{equation}
  \int\frac{dz}V\ g_3(x,y,z)=g_2(x,y)
  \label{g3g2}
\end{equation}
\begin{equation}
  \int\frac{dz}V\frac{dt}V\ g_4(x,y,z,t)=g_2(x,y)
  \label{g4g2}
\end{equation}
(all of which follow from\-~(\ref{grec})) to compute $u_i$ and $f_i$.
\bigskip

\indent
In the translation invariant case, the factorization assumption leads to an equation for $g_2$ alone, as $g_1$ is constant.
When translation invariance is broken, $g_1$ is no longer constant, and the Simplified approach consists in two coupled equations for $g_1$ and $g_2$.
We formulate these in terms of $g_1$ and $u_2$, with
\begin{equation}
  g_2(x,y)=:g_1(x)g_1(y)(1-u_2(x,y))
  .
\end{equation}
\bigskip

\theo{Theorem}\label{theo:simple}
  If $g_i$ satisfies Assumption\-~\ref{assum:factorization}, the eigenvalue equation\-~(\ref{eigval}) and\-~(\ref{g11})-(\ref{g4g2}), then $g_1$ and $u_2$ satisfy the two coupled equations
  \begin{equation}
    \left(
      -\frac\Delta 2
      +\left(\varpi-\left<\varpi\right>\right)
      +2\left(\mathcal E(x)-\left<\mathcal E(y)\right>\right)
      +\frac12\left(\bar A(x)-\left<\bar A\right>-\bar C(x)\right)
    \right)g_1(x)
    +\Sigma_1(x)
    =0
    \label{compleq_g1}
  \end{equation}
  and
  \begin{equation}
    \begin{largearray}
      \left(-\frac12(\Delta_x+\Delta_y)+v(x,y)-2\rho \bar K(x,y)+\rho^2\bar L(x,y)+\bar R_2(x,y)\right)
      g_1(x)g_1(y)(1-u_2(x,y))
      +\\\hfill+
      \Sigma_2(x,y)=0
      \label{compleq_g2}
    \end{largearray}
  \end{equation}
  where
  \begin{equation}
    \left<f\right>:=\int\frac{dy}V\ g_1(y)f(y)
    ,\quad
    \left<\varpi\right>\equiv \int\frac{dy}V\ \varpi g_1(y)
    \label{avgdef}
  \end{equation}
  \begin{equation}
    \bar S(x,y):=v(x,y)(1-u_2(x,y))
    ,\quad
    f_1\bar\ast f_2(x,y):=\int dz\ g_1(z)f_1(x,z)f_2(z,y)
  \end{equation}
  \begin{equation}
    \mathcal E(x):=
    \frac\rho2\int dy\ g_1(y)\bar S(x,y)
    ,\quad
    \bar A(x):=
    \rho^2\bar S\bar\ast u_2\bar\ast u_2(x,x)
    \label{EA}
  \end{equation}
  \begin{equation}
    \bar C(x):=
    2\rho^2\int dz\ g_1(z)u_2\bar\ast\bar S(x,z)
    +2\rho\int dy\ \varpi_y(g_1(y)u_2(x,y))
    .
    \label{C}
  \end{equation}
  \begin{equation}
    \bar K(x,y)
    :=
    \bar S\bar\ast u_2(x,y)
  \end{equation}
  \begin{equation}
    \begin{largearray}
      \bar L(x,y)
      :=
      \bar S\bar\ast u_2\bar\ast u_2(x,y)
      -2u_2\bar\ast(u_2(u_2\bar\ast\bar S))(x,y)
      +\\\hfill+
      \frac12\int dzdt\ g_1(z)g_1(t)\bar S(z,t) u_2(x,z)u_2(x,t)u_2(y,z)u_2(y,t)
    \end{largearray}
  \end{equation}
  \begin{equation}
    \begin{array}{r@{\ }>\displaystyle l}
      \bar R_2(x,y)
      =&
      2\left(\mathcal E(x)+\mathcal E(y)-2\left<\mathcal E\right>\right)
      +\left(\varpi_x+\varpi_y-2\left<\varpi\right>\right)
      +\\[0.3cm]&+
      \frac12\left(\bar A(x)+\bar A(y)-2\left<\bar A\right>-\bar C(x)-\bar C(y)\right)
      +2\rho u_2\bar\ast\left(u_2(\mathcal E-\left<\mathcal E\right>)\right)
      +\\[0.3cm]&+
      \rho\int dz\ \varpi_z(g_1(z)u_2(x,z)u_2(y,z))
      -
      \rho u_2\bar\ast u_2\left<\varpi\right>
      \label{R}
    \end{array}
  \end{equation}
  in which $\varpi_x$ is the action of $\varpi$ on the $x$-variable, and similarly for $\varpi_y$
  and
  \begin{equation}
    \Sigma_i\mathop{\longrightarrow}_{V\to\infty}0
  \end{equation}
  pointwise.
  Furthermore, the prediction for the energy per particle is
  \begin{equation}
    e:=\left<\mathcal E\right>+\left<\varpi\right>+\Sigma_0
    \label{simplen}
  \end{equation}
  where $\Sigma_0\to0$ as $V\to\infty$.
\endtheo
\bigskip

This theorem is proved in Section\-~\ref{sec:simple}.
\bigskip

\indent
Let us compare this to the equation for $u$ in the Simplified approach in the translation invariant case\-~\cite[(5)]{CHe21}, \cite[(3.15)]{Ja22}:
\begin{equation}
  -\Delta u(x)
  =
  (1-u(x))\left(v(x)-2\rho K(x)+\rho^2 L(x)\right)
  \label{compleq}
\end{equation}
\begin{equation}
  K:=
  u\ast S
  ,\quad
  S(y):=(1-u(y))v(y)
  \label{K}
\end{equation}
\nopagebreakaftereq
\begin{equation}
  L:=
  u\ast u\ast S
  -2u\ast(u(u\ast S))
  +\frac12
  \int dydz\ u(y)u(z-x)u(z)u(y-x)S(z-y)
  .
  \label{L}
\end{equation}
We will prove that these follow from Theorem\-~\ref{theo:simple}:
\bigskip

\theoname{Corollary}{Translation invariant case}\label{cor:check}
  In the translation invariant case $v(x,y)\equiv v(x-y)$ and $\varpi=0$ with periodic boundary conditions, if\-~(\ref{compleq_g1})-(\ref{compleq_g1}) has a unique translation invariant solution, then (\ref{compleq_g2}) reduces to\-~(\ref{compleq}) in the thermodynamic limit.
\endtheo
\bigskip

\indent
The idea of the proof is quite straightforward.
Equation\-~(\ref{compleq_g2}) is very similar to\-~(\ref{compleq}), but for the addition of the extra term $\bar R_2$.
An inspection of\-~(\ref{R}) shows that the terms in $\bar R_2$ are mostly of the form $f-\left<f\right>$, which vanish in the translation invariant case, and terms involving $\varpi$, which is set to 0 in the translation invariant case.
The only remaining extra term is $\bar C(x)+\bar C(y)$, which we will show vanishes in the translation invariant case due to the identity\-~(\ref{g2g1}).
\bigskip

\indent
Theorem\-~\ref{theo:simple} is quite general, and can be used to study a trapped Bose gas, in which there is an external potential $v_0$.
In this case, $\varpi$ is a multiplication operator by $v_0$.
A natural approach is to scale $v_0$ with the volume: $v_0(x)=\bar v_0(V^{-1/3}x)$ in such a way that the size of the trap grows as $V\to\infty$, thus ensuring a finite local density in the thermodynamic limit.
Following the ideas of Gross and Pitaevskii\-~\cite{Gr61,Pi61}, we would then expect to find that\-~(\ref{compleq_g1}) and\-~(\ref{compleq_g2}) decouple, and that (\ref{compleq_g2}) reduces to the translation invariant equation\-~(\ref{compleq}), with a density that is modulated over the trap.
However, the presence of $\bar R_2$ in\-~(\ref{compleq_g2}) and $\bar C$ in\-~(\ref{compleq_g1}) breaks this picture.
Further investigation of this question is warranted.

\subsection{The momentum distribution}\label{sec:Nk}
\indent
The momentum distribution for the Bose gas is defined as
\begin{equation}
  \mathcal M^{(\mathrm{Exact})}(k):=\frac1N\sum_{i=1}^N\left<\psi_0\right|P_i\left|\psi_0\right>
  \label{Mdef}
\end{equation}
where
\begin{equation}
  \varpi f:=\epsilon| e^{ikx}\big>\big< e^{ikx}|f
  \equiv
  \epsilon e^{ikx}\int dy\ e^{-iky}f(y)
  \label{varpiNk}
\end{equation}
and $P_i$ is defined as in\-~(\ref{Ppi}):
\begin{equation}
  P_i\psi(x_1,\cdots,x_N)= \epsilon e^{ikx_i}\int dy_y\ e^{iky_i}\psi(x_1,\cdots,x_{i-1},y_i,x_{i+1},\cdots,x_N)
\end{equation}
Equivalently,
\begin{equation}
  \mathcal M^{(\mathrm{Exact})}(k)=\frac\partial{\partial\epsilon}\left. \frac{E_0}N\right|_{\epsilon=0}
\end{equation}
where $E_0$ is the energy in\-~(\ref{eigval}) for the Hamiltonian\-~(\ref{ham}).
Using the Simplified approach, we do not have access to the ground state wavefunction, so we cannot compute $\mathcal M$ using\-~(\ref{Mdef}).
Instead, we use the Hellmann-Feynman theorem, which consists in adding $\sum_iP_i$ to the Hamiltonian.
However, doing so breaks the translational symmetry.
This is why Theorem\-~\ref{theo:simple} is needed to compute the momentum distribution.
(A similar computation was done in\-~\cite{CHe21}, but, there, the derivation of the momentum distribution for the Simplified approach was taken for granted.)
\bigskip

\indent
By Theorem\-~\ref{theo:simple}, and, in particular, (\ref{simplen}), we obtain a natural definition of the prediction of the Simplified approach for the momentum distribution:
\begin{equation}
  \mathcal M(k):=\frac{\partial}{\partial\epsilon}\left.\left(\left<\mathcal E\right>+\left<\varpi\right>\right)\right|_{\epsilon=0}
  .
\end{equation}
\bigskip

\theoname{Theorem}{Momentum distribution}\label{theo:Nk}
  Under the assumptions of Theorem\-~\ref{theo:simple}, using periodic boundary conditions, if $v$ is translation invariant and $\varpi=0$, then, if $k\neq 0$, in the thermodynamic limit,
  \begin{equation}
    \mathcal M(k)=\frac{\partial}{\partial\epsilon}\left.\frac\rho2\int dx\ (1-u(x))v(x)\right|_{\epsilon=0}
  \end{equation}
  where
  \begin{equation}
    -\Delta u(x)=(1-u(x))v(x)-2\rho K(x)+\rho^2L(x)+\epsilon F(x)
  \end{equation}
  where $K$ and $L$ are those of the translation invariant Simplified approach\-~(\ref{K})-(\ref{L}) and
  \begin{equation}
    F(x):=-2\hat u(-k)\cos(kx)
    .
    \label{F}
  \end{equation}
\endtheo
\bigskip

\indent
We thus compute the momentum distribution.
To check that our prediction is plausible, we compare it to the Bogolyubov prediction, which can easily be derived from\-~\cite[Appendix\-~A]{LSe05}:
\begin{equation}
  \mathcal M^{(\mathrm{Bogolyubov})}(k)=-\frac1{2\rho}\left(1-\frac{k^2+2\rho\hat v(k)}{\sqrt{k^4+4k^2\rho\hat v(k)}}\right)
\end{equation}
(this can be obtained by differentiating\-~\cite[(A.26)]{LSe05} with respect to $\epsilon(k)$, which returns the number of particles in the state $e^{ikx}$, which we divide by $\rho$ to obtain the momentum distribution).
Actually, following the ideas of\-~\cite{LHY57}, we replace $\hat v$ by a so-called ``pseudopotential'', which consists in replacing $v$ by a Dirac delta function, while preserving the scattering length:
\begin{equation}
  \hat v(k)=4\pi a
\end{equation}
where the scattering length $a$ is defined in\-~\cite[Appendix\-~C]{LSe05}.
Thus,
\begin{equation}
  \mathcal M^{(\mathrm{Bogolyubov})}(k)=-\frac1{2\rho}\left(1-\frac{k^2+8\pi\rho a}{\sqrt{k^4+16\pi k^2\rho a}}\right)
  .
  \label{Mbog}
\end{equation}
\bigskip

\indent
We prove that, for the Simple Equation, as $\rho\to0$, the prediction for the momentum distribution coincides with Bogolyubov's, for $|k|\lesssim\sqrt{\rho a}$.
The length scale $1/\sqrt{\rho a}$ is called the {\it healing length}, and is the distance at which pairs of particles correlate\-~\cite{FS20}.
It is reasonable to expect the Bogolyubov approximation to break down beyond this length scale.
\bigskip

\indent
The momentum distribution for the Simple equation, following the prescription detailed in\-~\cite{CJL20,CJL21,CHe21,Ja22}, is defined as
\begin{equation}
  \mathcal M^{(\mathrm{simpleq})}(k)=\frac{\partial}{\partial\epsilon}\left.\frac\rho2\int dx\ (1-u(x))v(x)\right|_{\epsilon=0}
  \label{M_simpleq}
\end{equation}
where\-~\cite[(1.1)-(1.2)]{CJL20}
\begin{equation}
  -\Delta u(x)=(1-u(x))v(x)-4eu+2\rho e u\ast u+\epsilon F(x)
  ,\quad
  e:=\frac\rho2\int dx\ (1-u(x))v(x)
  \label{simpleq}
\end{equation}
where $F$ was defined in\-~(\ref{F}).
\bigskip

\theo{Theorem}\label{theo:Nk_bog}
  Assume that $v$ is translation and rotation invariant ($v(x,y)\equiv v(|x-y|)$), and consider periodic boundary conditions.
  We rescale $k$:
  \begin{equation}
    \kappa:=\frac{k}{2\sqrt e}
  \end{equation}
  we have, for all $\kappa\in\mathbb R^3$,
  \begin{equation}
    \lim_{e\to0}\rho\mathcal M^{(\mathrm{simpleq})}(2\sqrt e\kappa)
    =\lim_{e\to0}\rho\mathcal M^{(\mathrm{Bogolyubov})}(2\sqrt e\kappa)
    =-\frac12\left(1-\frac{\kappa^2+1}{\sqrt{(\kappa^2+1)^2-1}}\right)
    .
    \label{Msimpleqbog}
  \end{equation}
\endtheo
\bigskip

\indent
The rotation invariance of $v$ is presumably not necessary.
However, the proof of this theorem is based on\-~\cite{CJL21}, where rotational symmetry was assumed for convenience.

\section{The Simplified approach without translation invariance, proof of Theorem \expandonce{\ref{theo:simple}}}\label{sec:simple}

\subsection{Factorization}
\indent
We will first compute $f_i$ and $u_i$ in Assumption\-~\ref{assum:factorization}.
\bigskip

\subsubsection{Factorization of $g_2$}
\indent
We start by considering $g_2$.
\bigskip

\theo{Lemma}\label{lemma:g2}
  Assumption\-~\ref{assum:factorization} with $i=2$ and\-~(\ref{g11})-(\ref{g2g1}) imply that
  \begin{equation}
    g_2(x,y)=g_1(x)g_1(y)(1-u(x,y))(1+O(V^{-2}))
    .
  \end{equation}
\endtheo

\indent\underline{Proof}:
  Assumption\-~\ref{assum:factorization} implies
  \begin{equation}
    g_2(x,y)=f_2(x)f_2(y)(1-u_2(x,y))
    .
  \end{equation}
  and by\-~(\ref{g2g1}),
  \begin{equation}
    g_1(x)=f_2(x)\int \frac{dy}V f_2(y)(1-u_2(x,y))
    .
    \label{g1_fact}
  \end{equation}
  \bigskip

  \point
  Let us first take an expansion to order $V^{-1}$.
  By~\-(\ref{assum_bound})
  \begin{equation}
    \int\frac{dy}V\ f_2(y)u_2(x,y)=O(V^{-1})
  \end{equation}
  and so
  \begin{equation}
    g_1(x)=f_2(x)\left(\int\frac{dy}V f_2(y)+O(V^{-1})\right)
    .
    \label{g1f2}
  \end{equation}
  Applying $\int\frac{dx}V\cdot$ to both sides of\-~(\ref{g1f2}), we find that
  \begin{equation}
    \int\frac{dy}Vf_2(y)=1+O(V^{-1})
  \end{equation}
  so\-~(\ref{g1f2}) yields
  \begin{equation}
    f_2(x)=g_1(x)(1+O(V^{-1}))
    .
    \label{f1V1}
  \end{equation}
  \bigskip

  \point
  We now push the expansion to order $V^{-2}$.
  Inserting\-~(\ref{f1V1}) into\-~(\ref{g1_fact}),
  \begin{equation}
    g_1(x)=f_2(x)\int\frac{dy}V\ f_2(y)-g_1(x)\left(\int\frac{dy}V\ g_1(y)u_2(x,y)+O(V^{-2})\right)
    .
  \end{equation}
  However, by\-~(\ref{g2g1}),
  \begin{equation}
    g_1(x)\int\frac{dy}V\ g_1(y)(1-u_2(x,y))=g_1(x)
  \end{equation}
  so, by\-~(\ref{g11}),
  \begin{equation}
    \int dy\ g_1(y)u_2(x,y)=0
    \label{intu0}
  \end{equation}
  and
  \begin{equation}
    g_1(x)(1+O(V^{-2}))=f_2(x)\int\frac{dy}V\ f_2(y)
    .
  \end{equation}
  Taking $\int\frac{dx}V\cdot$ on both sides, we find that
  \begin{equation}
    f_2(x)=g_1(x)(1+O(V^{-2}))
    .
  \end{equation}
\qed
\bigskip

{\bf Remark}:
Note that this proof can easily be generalized to show that $f_2=g_1(1+O(V^{-n}))$ for any $n$.

\subsubsection{Factorization of $g_3$}
\indent
We now turn to $g_3$.
\bigskip

\theo{Lemma}\label{lemma:g3}
  Assumption\-~\ref{assum:factorization} with $i=2,3$ and\-~(\ref{g11})-(\ref{g3g2}) imply that
  \begin{equation}
    g_3(x,y,z)=g_1(x)g_1(y)g_1(z)(1-u_3(x,y))(1-u_3(x,z))(1-u_3(y,z))(1+O(V^{-2}))
  \end{equation}
  with
  \begin{equation}
    u_3(x,y):=u_2(x,y)+\frac{w_3(x,y)}V
    \label{u3}
  \end{equation}
  \begin{equation}
    w_3(x,y):=(1-u_2(x,y))\int dz\ g_1(z)u_2(x,z)u_2(y,z)
    .
    \label{w3}
  \end{equation}
\endtheo
\bigskip

\indent\underline{Proof}:
  Using\-~(\ref{g3g2}) in\-~(\ref{g_factorized}),
  \begin{equation}
    g_2(x_1,x_2)=W_3(x_1,x_2)
    \int \frac{dx_3}V\ W_3(x_1,x_3)W_3(x_2,x_3)
    .
    \label{g2_factor_inproof}
  \end{equation}
  \bigskip

  \point
  We first expand to order $V^{-1}$.
  By\-~(\ref{assum_bound}),
  \begin{equation}
    \int\frac{dz}Vf_3^2(z)u_3(x,z)=O(V^{-1})
    \label{f3V1}
  \end{equation}
  so, by\-~(\ref{W_fact}),
  \begin{equation}
    g_2(x,y)=f_3^2(x)f_3^2(y)(1-u_3(x,y))
    \left(\int \frac{dz}V\ f_3^2(z)
    +O(V^{-1})\right)
    .
  \end{equation}
  By Lemma~\-\ref{lemma:g2},
  \begin{equation}
    g_1(x)g_1(y)(1-u_2(x,y))=f_3^2(x)f_3^2(y)(1-u_3(x,y))\left(\int\frac{dz}V\ f_3^2(z)+O(V^{-1})\right)
    .
  \end{equation}
  We take $\int\frac{dy}V\cdot$ on both sides of this equation.
  By\-~(\ref{intu0}) and\-~(\ref{f3V1}),
  \begin{equation}
    g_1(x)=f_3^2(x)\left(\left(\int\frac{dy}Vf_3^2(y))\right)^2+O(V^{-1})\right)
  \end{equation}
  and, integrating once more implies that $\int\frac{dy}Vf_3^2(y)=1+O(V^{-1})$.
  Therefore,
  \begin{equation}
    f_3^2(x)=g_1(x)(1+O(V^{-1}))
    \label{3fV}
  \end{equation}
  and
  \begin{equation}
    u_3(x,y)=u_2(x,y)(1+O(V^{-1}))
    .
    \label{3V}
  \end{equation}
  \bigskip

  \point
  We push the expansion to order $V^{-2}$: (\ref{g2_factor_inproof}) is
  \begin{equation}
    g_2(x,y)=f_3^2(x)f_3^2(y)(1-u_3(x,y))\int\frac{dz}{V}f_3^2(z)
    \left(
      1
      -u_3(x,z)-u_3(y,z)
      +u_3(x,z)u_3(y,z)
    \right)
    .
  \end{equation}
  By\-~(\ref{3fV})-(\ref{3V}) and Lemma\-~\ref{lemma:g2},
  \begin{equation}
    \begin{largearray}
      f_3^2(x)f_3^2(y)(1-u_3(x,y))\int\frac{dz}{V}f_3^2(z)
      =g_1(x)g_1(y)(1-u_2(x,y))
      \cdot\\[0.3cm]\hfill\cdot
      \left(1+\int\frac{dz}{V}\ (g_1(z)(u_2(x,z)+u_2(y,z)-u_2(x,z)u_2(y,z)))+O(V^{-2})\right)
      .
    \end{largearray}
  \end{equation}
  Therefore, by\-~(\ref{intu0}),
  \begin{equation}
    \begin{largearray}
      f_3^2(x)f_3^2(y)(1-u_3(x,y))\int\frac{dz}{V}f_3^2(z)=g_1(x)g_1(y)(1-u_2(x,y))
      \cdot\\\hfill\cdot
      \left(1-\int\frac{dz}{V}g_1(z)u_2(x,z)u_2(y,z)+O(V^{-2})\right)
      .
    \end{largearray}
  \end{equation}
  Now, let us apply $\int\frac{dy}V\cdot$ to both sides of the equation.
  Note that, by\-~(\ref{assum_bound}),
  \begin{equation}
    \int\frac{dy}V\ g_1(y)u_2(x,y)\int\frac{dz}Vg_1(z)u_2(x,z)u_2(y,z)=O(V^{-2})
    .
    \label{tech1}
  \end{equation}
  Furthermore, by\-~(\ref{intu0}),
  \begin{equation}
    \int \frac{dy}V\ g_1(y)u_2(x,y)=0
    ,\quad
    \int\frac{dy}V\ g_1(y)\int\frac{dz}V\ g_1(z)u_2(x,z)u_2(y,z)=0
  \end{equation}
  and by\-~(\ref{3fV}) and\-~(\ref{3V}),
  \begin{equation}
    \int\frac{dy}V\ f_3^2(y)u_3(x,y)=\int\frac{dy}V\ g_1(y)u_2(x,y)+O(V^{-2})=O(V^{-2})
    .
    \label{tech2}
  \end{equation}
  We are thus left with
  \begin{equation}
    f_3^2(x)\left(\int\frac{dy}V\ f_3^2(y)\right)^2
    =
    g_1(x)(1+O(V^{-2}))
    .
  \end{equation}
  Taking $\int\frac{dx}V\cdot$, we thus find that
  \begin{equation}
    \left(\int\frac{dx}V f_3^2(x)\right)^3=1+O(V^{-2})
  \end{equation}
  and
  \begin{equation}
    f_3^2(x)=g_1(x)(1+O(V^{-2}))
    .
  \end{equation}
  Therefore,
  \begin{equation}
    1-u_3(x,y)=(1-u_2(x,y))\left(1-\frac1V\int dz\ g_1(z)u_2(x,z)u_2(y,z)+O(V^{-2})\right)
    .
  \end{equation}
\qed

\subsubsection{Factorization of $g_4$}

\theo{Lemma}\label{lemma:g4}
  Assumption\-~\ref{assum:factorization} and\-~(\ref{g11})-(\ref{g4g2}) imply that
  \begin{equation}
    g_4(x_1,x_2,x_3,x_2)=
    g_1(x_1)g_1(x_2)g_1(x_3)g_1(x_4)
    \left(\prod_{i<j}(1-u_4(x_i,x_j))\right)
    (1+O(V^{-2}))
  \end{equation}
  with
  \begin{equation}
    u_4(x,y):=u_2(x,y)+\frac{2w_3(x,y)}V
    \label{u4}
  \end{equation}
  where $w_3$ is the same as in Lemma\-~\ref{lemma:g3}.
\endtheo
\bigskip

\indent\underline{Proof}:
  Using\-~(\ref{g4g2}) in\-~(\ref{g_factorized}),
  \begin{equation}
    g_2(x_1,g_2)=W_4(x_1,x_2)\int\frac{dx_3dx_4}{V^2}\ 
    W_4(x_1,x_3)
    W_4(x_1,x_4)
    W_4(x_2,x_3)
    W_4(x_2,x_4)
    W_4(x_3,x_4)
    .
  \end{equation}
  \bigskip

  \point
  We expand to order $V^{-1}$.
  By\-~(\ref{assum_bound}),
  \begin{equation}
    \int\frac{dz}Vf_4^3(z)u_4(x,z)=O(V^{-1})
    \label{f4V1}
  \end{equation}
  so by\-~(\ref{W_fact}),
  \begin{equation}
    g_2(x,y)=f_4^3(x)f_4^3(y)(1-u_4(x,y))\left(\int\frac{dzdt}{V^2}f_4^3(z)f_4^3(t)+O(V^{-1})\right)
    .
  \end{equation}
  By Lemma\-~\ref{lemma:g2},
  \begin{equation}
    g_1(x)g_1(y)(1-u_2(x,y))=
    f_4^3(x)f_4^3(y)(1-u_4(x,y))\left(\left(\int\frac{dz}{V}f_4^3(z)\right)^2+O(V^{-1})\right)
    .
  \end{equation}
  Applying $\int\frac{dy}V\cdot$ to both sides of the equation, using\-~(\ref{intu0}) and\-~(\ref{f4V1}),
  \begin{equation}
    g_1(x)=f_4(x)^3\left(\left(\int\frac{dy}V\ f_4^3(y)\right)^3+O(V^{-1})\right)
    .
  \end{equation}
  Integrating once more, we have $\int\frac{dy}Vf_4^3(z)=1+O(V^{-1})$ and
  \begin{equation}
    f_4^3(x)=g_1(x)(1+O(V^{-1}))
    .
    \label{4fV}
  \end{equation}
  Therefore,
  \begin{equation}
    u_4(x,y)=u_2(x,y)(1+O(V^{-1}))
    .
    \label{4V}
  \end{equation}
  \bigskip

  \point
  We push the expansion to order $V^{-2}$:
  by\-~(\ref{assum_bound}),
  \begin{equation}
    \int \frac{dzdt}{V^2}u_4(x,z)u_4(y,t)=O(V^{-2})
    ,\quad
    \int \frac{dzdt}{V^2}u_4(x,z)u_4(z,t)=O(V^{-2})
  \end{equation}
  \begin{equation}
    \int \frac{dzdt}{V^2}u_4(x,z)u_4(x,t)=O(V^{-2})
  \end{equation}
  so
  \begin{equation}
    \begin{largearray}
      g_2(x,y)=f_4^3(x)f_4^3(y)(1-u_4(x,y))
      \left(\int\frac{dzdt}{V^2}
	f_4^3(z)f_4^3(t)
	+\right.\\[0.5cm]\hfill\left.+
	\int\frac{dzdt}{V^2}
	g_1(z)g_1(t)(-2u_2(x,z)-2u_2(y,z)-u_2(z,t)+2u_2(x,z)u_2(y,z))
	+O(V^{-2})
      \right)
      .
    \end{largearray}
  \end{equation}
  By\-~(\ref{4fV}), (\ref{4V}), and Lemma\-~\ref{lemma:g2},
  \begin{equation}
    \begin{largearray}
      f_4^3(x)f_4^3(x)(1-u_4(x,y))\left(\int\frac{dz}V\ f_4^3(z)\right)^2
      =
      g_1(x)g_1(y)(1-u_2(x,y))
      \cdot\\[0.5cm]\hfill\cdot
      \left(1+
	\int\frac{dzdt}{V^2}\ g_1(z)g_1(t)(2u_2(x,z)+2u_2(y,z)+u_2(z,t)-2u_2(x,z)u_2(y,z))
	+O(V^{-2})
      \right)
      .
    \end{largearray}
  \end{equation}
  By~\-(\ref{intu0}),
  \begin{equation}
    \begin{largearray}
      f_4^3(x)f_4^3(y)(1-u_4(x,y))\left(\int\frac{dz}V\ f_4^3(z)\right)^2
      =\\[0.3cm]\hfill=
      g_1(x)g_1(y)(1-u_2(x,y))\left(1-2\int\frac{dz}{V}g_1(z)u_2(x,z)u_2(y,z)+O(V^{-2})\right)
      .
    \end{largearray}
  \end{equation}
  We apply $\int\frac{dy}V\cdot$ to both sides of the equation.
  By\-~(\ref{tech1})-(\ref{tech2}), we find
  \begin{equation}
    f_4^3(x)\left(\int\frac{dy}Vf_4^3(z)\right)^3=g_1(x)(1+O(V^{-2}))
    .
  \end{equation}
  Taking $\int\frac{dx}V\cdot$, we find that
  \begin{equation}
    f_4(x)=1+O(V^{-2})
  \end{equation}
  and
  \begin{equation}
    f_4^3(x)=g_1(x)(1+O(V^{-2}))
    .
  \end{equation}
  Therefore,
  \begin{equation}
    1-u_4(x,y)=(1-u_2(x,y))\left(1-\frac2V\int dz\ g_1(z)u_2(x,z)u_2(y,z)+O(V^{-2})\right)
    .
  \end{equation}
\qed

\subsection{Consequences of the factorization}

\point
We first rewrite\-~(\ref{eigval}) as a family of equations for $g_i$.
\bigskip

\subpoint
Integrating~\-(\ref{eigval}) with respect to $x_1,\cdots,x_N$, we find that
\begin{equation}
  E_0=
  G^{(2)}_0
  +F^{(1)}_0
  +B_0
  \label{E0}
\end{equation}
with
\begin{equation}
  G^{(2)}_0:=
  \frac{N(N-1)}{2V^2}\int dxdy\ v(x,y)g_2(x,y)
\end{equation}
\begin{equation}
  F^{(1)}_0:=
  \frac NV\int dx\ \varpi g_1(x)
\end{equation}
and $B_0$ is a boundary term:
\begin{equation}
  B_0=-\frac{N}{2V}\int dx\ \Delta g_1(x)
  .
\end{equation}
\bigskip

\subpoint
If, now, we integrate~\-(\ref{eigval}) with respect to $x_2,\cdots,x_N$, we find
\begin{equation}
  -\frac\Delta 2g_1(x)
  +\varpi g_1(x)
  +G^{(2)}_1(x)
  +G^{(3)}_1(x)
  +F^{(2)}_1(x)
  +B_1(x)
  =E_0g_1(x)
  \label{g1}
\end{equation}
with
\begin{equation}
  G^{(2)}_1(x):=\frac{N-1}V\int dy\ v(x,y)g_2(x,y)
\end{equation}
\begin{equation}
  G^{(3)}_1(x):=
  \frac{(N-1)(N-2)}{2V^2}\int dydz\ v(y,z)g_3(x,y,z)
\end{equation}
\begin{equation}
  F^{(2)}_1(x):=\frac{N-1}V\int dy\ \varpi_y g_2(x,y)
\end{equation}
in which we use the notation $\varpi_y$ to indicate that $\varpi$ applies to $y\mapsto g_2(x,y)$,
and $B_1$ is a boundary term
\begin{equation}
  B_1(x):=-\frac{N-1}{2V}\int dy\ \Delta_y g_2(x,y)
  .
\end{equation}
\bigskip

\subpoint
If we integrate with respect to $x_3,\cdots,x_N$, we find
\begin{equation}
  \begin{largearray}
    -\frac12(\Delta_x+\Delta_y)g_2(x,y)
    +v(x,y)g_2(x,y)
    +(\varpi_y+\varpi_x)g_2(x,y)
    +\\\hfill+
    G^{(3)}_2(x,y)
    +G^{(4)}_2(x,y)
    +F^{(3)}_2(x,y)
    +B_2(x,y)
    =E_0g_2(x,y)
    \label{g2}
  \end{largearray}
\end{equation}
where, here again, $\varpi_y$ indicates that $\varpi$ applies to the $y$-degree of freedom, whereas $\varpi_x$ applies to $x$,
with
\begin{equation}
  G^{(3)}_2(x,y):=
  \frac{N-2}V\int dz\ (v(x,z)+v(y,z))g_3(x,y,z)
\end{equation}
\begin{equation}
  G^{(4)}_2(x,y):=
  \frac{(N-2)(N-3)}{2V^2}\int dzdt\ v(z,t)g_4(x,y,z,t)
\end{equation}
\begin{equation}
  F^{(3)}_2(x,y):=
  \frac{N-2}V\int dz\ \varpi_z g_3(x,y,z)
\end{equation}
and $B_2$ is a boundary term
\begin{equation}
  B_2(x):=-\frac{N-2}{2V}\int dz\ \Delta_z g_3(x,y,z)
  .
\end{equation}

\point
We rewrite\-~(\ref{E0}), (\ref{g1}) and~\-(\ref{g2}) using Lemmas\-~\ref{lemma:g2}, \ref{lemma:g3} and\-~\ref{lemma:g4}.
\bigskip

\subpoint We start with\-~(\ref{E0}): by\-~(\ref{intv}) and Lemma\-~\ref{lemma:g2},
\begin{equation}
  G_0^{(2)}=
  \frac{N(N-1)}{2V^2}\int dxdy\ v(x,y)g_1(x)g_1(y)(1-u_2(x,y))+O(V^{-1})
\end{equation}
so
\begin{equation}
  E_0=
  \frac{N(N-1)}{2V^2}\int dxdy\ v(x,y)g_1(x)g_1(y)(1-u_2(x,y))
  +
  \frac NV\int dx\ \varpi g_1(x)
  +B_0
  +O(V^{-1})
  .
\end{equation}
\bigskip

\subpoint We now turn to\-~(\ref{g1}): by\-~(\ref{intv}) and Lemma\-~\ref{lemma:g2},
\begin{equation}
  G_1^{(2)}(x)=\frac{N}Vg_1(x)\left(\int dy\ v(x,y)g_1(y)(1-u_2(x,y))+O(V^{-2})\right)
\end{equation}
and by Lemma\-~\ref{lemma:g3},
\begin{equation}
  \begin{largearray}
    G_1^{(3)}(x)=
    g_1(x)\left(\frac{N^2}{2V^2}\int dydz\ v(y,z)g_1(y)g_1(z)(1-u_2(x,y))(1-u_2(x,z))(1-u_3(y,z))
    -\right.\\\hfill\left.-
    \frac{3N}{2V^2}\int dydz\ v(y,z)g_1(y)g_1(z)(1-u_2(y,z))
    +O(V^{-1})\right)
  \end{largearray}
\end{equation}
(we used\-~(\ref{u3}) to write $u_3=u_2+O(V^{-1})$; this works fine for $u_3(x,y)$ and $u_3(x,z)$ because the integrals over $y$ and $z$ are controlled by $v(y,z)w_3(x,y)$ and $v(y,z)w_3(x,z)$ using\-~(\ref{intv}) and\-~(\ref{assum_bound}); in the first term, it does not work for $u_3(y,z)$, as $v(y,z)w_3(y,z)$ can only control one of the integrals, and not both; the second term has an extra $V^{-1}$ that lets us replace $u_3$ by $u_2$)
and by\-~(\ref{assum_bound}) and\-~(\ref{bound_varpi}),
\begin{equation}
  F_1^{(2)}(x)=
  g_1(x)\left(\frac NV\int dy\ \varpi_y(g_1(y)(1-u_2(x,y)))
  -\frac 1V\int dy\ \varpi g_1(y)
  +O(V^{-1})
  \right)
  .
\end{equation}
The first term in $G_1^{(3)}$ is of order $V$:
\begin{equation}
  \begin{largearray}
    \frac{N^2}{2V^2}\int dydz\ v(y,z)g_1(y)g_1(z)(1-u_2(x,y))(1-u_2(x,z))(1-u_3(y,z))
    =\\[0.3cm]\indent=
    \frac{N^2}{2V^2}\int dydz\ v(y,z)g_1(y)g_1(z)(1-u_2(y,z))
    -
    \frac{N^2}{2V^3}\int dydz\ v(y,z)g_1(y)g_1(z)w_3(y,z)
    +\\[0.5cm]\hfill+
    \frac{N^2}{2V^2}\int dydz\ v(y,z)g_1(y)g_1(z)(1-u_2(y,z))(-u_2(x,y)-u_2(x,z)+u_2(x,y)u_2(x,z))
    +O(V^{-1})
  \end{largearray}
\end{equation}
in which the only term of order $V$ is the first one, and is equal to the first term of order $V$ in $E_0$, and thus cancels out.
There is a similar cancellation between the second term of order $V$ in $F_1^{(2)}$ and $E_0$. All in all,
\begin{equation}
  \left(
    -\frac\Delta 2
    +\varpi
    +\bar G^{(2)}_1(x)
    +\bar G^{(3)}_1(x)
    +\bar F^{(2)}_1(x)
    +\bar E_0
    -B_0
  \right)
  g_1(x)
  +B_1(x)
  =g_1(x)O(V^{-1})
  \label{g1bar}
\end{equation}
with, recalling $\rho:=N/V$,
\begin{equation}
  \bar G_1^{(2)}(x):=\rho\int dy\ v(x,y)g_1(y)(1-u_2(x,y))
\end{equation}
and using\-~(\ref{w3}),
\begin{equation}
  \begin{largearray}
    \bar G_1^{(3)}(x):=
    -\frac\rho2\int \frac{dydz}V\ v(y,z)g_1(y)g_1(z)(1-u_2(y,z))\left(3+\rho \int dt\ g_1(t)u_2(y,t)u_2(z,t)\right)
    +\\[0.5cm]\hfill+
    \frac{\rho^2}2\int dydz\ v(y,z)g_1(y)g_1(z)(1-u_2(y,z))(-u_2(x,y)-u_2(x,z)+u_2(x,y)u_2(x,z))
  \end{largearray}
\end{equation}
\begin{equation}
  \bar F_1^{(2)}(x):=
  -\rho\int dy\ \varpi_y(g_1(y)u_2(x,y))
  -\int \frac{dy}V\ \varpi g_1(y)
\end{equation}
\begin{equation}
  \bar E_0:=
  \frac\rho2\int \frac{dxdy}V\ v(x,y)g_1(x)g_1(y)(1-u_2(x,y))
  .
\end{equation}
Rewriting this using\-~(\ref{avgdef})-(\ref{C}), we find\-~(\ref{compleq_g1}) with
\begin{equation}
  \Sigma_1(x):=B_1(x)-B_0g_1(x)+O(V^{-1})
  .
\end{equation}
\bigskip

\subpoint Finally, we rewrite (\ref{g2}): by\-~(\ref{intv}) and Lemma\-~\ref{lemma:g3},
\begin{equation}
  \begin{largearray}
    G_2^{(3)}(x,y)=
    \frac NVg_1(x)g_1(y)(1-u_2(x,y))
    \cdot\\\hfill\cdot
    \left(\int dz\ (v(x,z)+v(y,z))g_1(z)(1-u_2(x,z))(1-u_2(y,z))+O(V^{-1})\right)
  \end{largearray}
\end{equation}
and by Lemma\-~\ref{lemma:g4},
\begin{equation}
  \begin{largearray}
    G^{(4)}_2(x,y)=
    g_1(x)g_1(y)\left(\frac{N^2}{2V^2}(1-u_4(x,y))\int dzdt\ v(z,t)g_1(z)g_1(t)(1-u_4(z,t))\Pi(x,y,z,t)
    -\right.\\\hfill\left.-
    \frac{5N}{2V^2}(1-u_2(x,y))\int dzdt\ v(z,t)g_1(z)g_1(t)(1-u_2(z,t))
    +O(V^{-1})
    \right)
  \end{largearray}
\end{equation}
\begin{equation}
  \Pi(x,y,z,t):=
  (1-u_2(x,z))(1-u_2(x,t))(1-u_2(y,z))(1-u_2(y,t))
  \label{Pi}
\end{equation}
and by\-~(\ref{assum_bound}) and\-~(\ref{bound_varpi}),
\begin{equation}
  \begin{largearray}
    F^{(3)}_2(x,y)=
    g_1(x)g_1(y)\left(
    \frac NV(1-u_3(x,y))\int dz\ \varpi_z(g_1(z)(1-u_2(x,z))(1-u_2(y,z)))
    -\right.\\\hfill\left.-
    \frac2V(1-u_2(x,y))\int dz\ \varpi g_1(z)
    +O(V^{-1})
    \right)
    .
  \end{largearray}
\end{equation}

\vfill
\eject

The first term in $G_2^{(4)}$ is of order $V$: by\-~(\ref{u4}),
\begin{equation}
  \begin{largearray}
    \frac{N^2}{2V^2}(1-u_4(x,y))\int dzdt\ v(z,t)g_1(z)g_1(t)(1-u_4(z,t))
    \Pi(x,y,z,y)
    =\\[0.5cm]\indent=
    \frac{N^2}{2V^2}(1-u_2(x,y))\int dzdt\ v(z,t)g_1(z)g_1(t)(1-u_2(z,t))
    -\\[0.5cm]\indent-
    \frac{N^2}{V^3}w_3(x,y)\int dzdt\ v(z,t)g_1(z)g_1(y)(1-u_2(z,t))
    -\\[0.5cm]\indent-
    \frac{N^2}{V^3}(1-u_2(x,y))\int dzdt\ v(z,t)g_1(z)g_1(t)w_3(z,t)
    +\\[0.5cm]\indent+
    \frac{N^2}{2V^2}(1-u_2(x,y))\int dzdt\ v(z,t)g_1(z)g_1(t)(1-u_2(z,t))
    \left(\Pi(x,y,z,t)-1\right)
    +O(V^{-1})
  \end{largearray}
\end{equation}
in which the only term of order $V$ is the first one, and is equal to the term of order $V$ in $E_0$, and thus cancels out.
There is a similar cancellation between the term of order $V$ in $F_2^{(3)}$ and $E_0$.
All in all,
\begin{equation}
  \begin{largearray}
    \left(
      -\frac12(\Delta_x+\Delta_y)
      +v(x,y)
      +\varpi_x+\varpi_y
      +\bar G^{(3)}_2(x,y)
      +\bar G^{(4)}_2(x,y)
      +\bar F^{(3)}_2(x,y)
      +\bar E_0
      -B_0
    \right)
    \cdot\\\hfill\cdot
    g_1(x)g_1(y)(1-u_2(x,y))
    +B_2(x,y)
    =
    g_1(x)g_1(y)O(V^{-1})
  \end{largearray}
  \label{g2bar}
\end{equation}
with
\begin{equation}
  \bar G_2^{(3)}(x,y):=
  \rho\int dz\ (v(x,z)+v(y,z))g_1(z)(1-u_2(x,z))(1-u_2(y,z))
\end{equation}
and by\-~(\ref{w3}),
\begin{equation}
  \begin{array}{r@{\ }>\displaystyle l}
    \bar G_2^{(4)}(x,y)
    :=&
    -\frac\rho2\left(5+2\rho \int dr\ g_1(r)u_2(x,r)u_2(y,r)\right)\int \frac{dzdt}V\ v(z,t)g_1(z)g_1(t)(1-u_2(z,t))
    -\\[0.3cm]&-
    \rho^2\int \frac{dzdt}V\ v(z,t)g_1(z)g_1(t)(1-u_2(z,t))\int dr\ g_1(r)u_2(z,r)u_2(t,r)
    +\\&+
    \frac{\rho^2}2\int dzdt\ v(z,t)g_1(z)g_1(t)(1-u_2(z,t))
    \left(\Pi(x,y,z,t)-1\right)
  \end{array}
\end{equation}
\begin{equation}
  \begin{largearray}
    \bar F^{(3)}_2(x,y):=
    \rho\int dz\ \varpi_z(g_1(z)(-u_2(x,z)-u_2(y,z)+u_2(x,z)u_2(y,z)))
    -\\\hfill-
    \left(2+\rho\int dr\ g_1(r)u_2(x,r)u_2(y,r)\right)\int \frac{dz}V\ \varpi g_1(z)
  \end{largearray}
\end{equation}
\begin{equation}
  \bar E_0=
  \frac\rho2\int \frac{dxdy}V\ v(x,y)g_1(x)g_1(y)(1-u_2(x,y))
  .
\end{equation}
\bigskip

\subpoint
Expanding out $\Pi$, see\-~(\ref{Pi}), we find\-~(\ref{compleq_g2}) with
\begin{equation}
  \begin{array}{r@{\ }>\displaystyle l}
    \bar R_2(x,y)
    :=&
    \rho\int dz\ g_1(z)\left(
      \bar S(x,z)+\bar S(y,z)
      -2\int \frac{dt}V\ g_1(t)\bar S(t,z)
    \right)
    +\\[0.3cm]&+
    \frac{\rho^2}2\left(
      \bar S\bar\ast u_2\bar\ast u_2(x,x)+\bar S\bar\ast u_2\bar\ast u_2(y,y)
      -2\int \frac{dt}V\ g_1(t)\bar S\bar\ast u_2\bar\ast u_2(t,t)
    \right)
    +\\[0.3cm]&+
    \rho^2\int dzdt\ g_1(z)g_1(t)u_2(x,z)u_2(y,z)\left(
      \bar S(z,t)
      -\int\frac{dr}V\ g_1(r)\bar S(z,r)
    \right)
    -\\[0.3cm]&-
    \rho^2\int dt\ g_1(t)(\bar S\bar\ast u_2(x,t)+\bar S\bar\ast u_2(y,t))
    +\bar F_2^{(3)}(x,y)
    +\varpi_x+\varpi_y
  \end{array}
  \label{R1}
\end{equation}
and
\begin{equation}
  \Sigma_2(x,y):=B_2(x,y)-B_0g_1(x)g_1(y)(1-u_2(x,y))+O(V^{-1})
  .
\end{equation}
Using\-~(\ref{EA}) and\-~(\ref{C}), (\ref{R1}) becomes\-~(\ref{R}).
\bigskip

\point
Finally, (\ref{simplen}) follows from\-~(\ref{E0}) with
\begin{equation}
  \Sigma_0:=B_0+O(V^{-1})
  .
\end{equation}

\qed

\subsection{Sanity check, proof of Corollary \expandonce{\ref{cor:check}}}\label{sec:trsl_inv}
\indent
Assuming the translation invariance of the solution, $g_1(x)$ is constant.
By\-~(\ref{g11}),
\begin{equation}
  g_1(x)=1
  .
  \label{g1const}
\end{equation}
Furthermore, $\varpi\equiv 0$.
We then have
\begin{equation}
  \bar S(x,y)=S(x-y)
  ,\quad
  \bar K(x,y)=K(x-y)
  ,\quad
  \bar L(x,y)=L(x-y)
\end{equation}
(see\-~(\ref{K})-(\ref{L})).
Furthermore,
\begin{equation}
  \mathcal E(x)\equiv \mathcal E(y)\equiv\left<\mathcal E\right>=\frac\rho2\int dy\ S(y)
\end{equation}
\begin{equation}
  \bar A(x)\equiv\bar A(y)\equiv\left<\bar A\right>=\rho^2 S\ast u\ast u(0)
\end{equation}
\begin{equation}
  \bar C(x)\equiv \bar C_2(y)
  =2\rho^2\int dz\ u(z)\int dt\ S(t)
\end{equation}
which vanishes by\-~(\ref{g2g1}).
Thus,
\begin{equation}
  \bar R_2(x,y)\equiv0
  .
\end{equation}
We conclude by taking the thermodynamic limit.
\qed

\section{The momentum distribution}

\subsection{Computation of the momentum distribution, proof of Theorem \expandonce{\ref{theo:Nk}}}\label{sec:Nk_proof}
\indent
We use Theorem\-~\ref{theo:simple} with $\varpi$ as in\-~(\ref{varpiNk}).
Note that, by\-~(\ref{varpiNk}),
\begin{equation}
  \int dx\ \varpi f(x)=0
\end{equation}
which trivially satisfies\-~(\ref{bound_varpi}).
\bigskip

\point
We change variables in\-~(\ref{compleq_g2}) to
\begin{equation}
  \xi=\frac{x+y}2
  ,\quad
  \zeta=x-y
\end{equation}
and find
\begin{equation}
  \begin{largearray}
    \left(
      -\frac14\Delta_\xi-\Delta_\zeta+v(\zeta)
      -2\rho\bar K(\xi+{\textstyle\frac\zeta 2},\xi-{\textstyle\frac\zeta 2})
      +\rho^2\bar L(\xi+{\textstyle\frac\zeta 2},\xi-{\textstyle\frac\zeta 2})
      +\bar R_2(\xi+{\textstyle\frac\zeta 2},\xi-{\textstyle\frac\zeta 2})
    \right)
    \cdot\\\hfill\cdot
    g_1(\xi+{\textstyle\frac\zeta 2})g_1(\xi-{\textstyle\frac\zeta 2})
    (1-u_2(\xi+{\textstyle\frac\zeta 2},\xi-{\textstyle\frac\zeta 2}))
    =-\Sigma_2
    .
    \label{g2_xi}
  \end{largearray}
\end{equation}
In addition, by\-~(\ref{simplen}),
\begin{equation}
  e=\frac\rho2\int \frac{d\xi d\zeta}V\ g_1(\xi+{\textstyle\frac\zeta 2})g_1(\xi-{\textstyle\frac\zeta 2})v(\zeta)(1-u_2(\xi+{\textstyle\frac\zeta 2},\xi-{\textstyle\frac\zeta 2}))
  +\int\frac{dx}V\ \varpi g_1(x)
  +\Sigma_1
  .
  \label{Nken}
\end{equation}
We expand in powers of $\epsilon$:
\begin{equation}
  g_1(x)=1+\epsilon g_1^{(1)}(x)+O(\epsilon^2)
  ,\quad
  u_2(\xi+{\textstyle\frac\zeta2},\xi-{\textstyle\frac\zeta 2})=u_2^{(0)}(\zeta)+\epsilon u_2^{(1)}(\xi+{\textstyle\frac\zeta2},\xi-{\textstyle\frac\zeta 2})+O(\epsilon^2)
\end{equation}
in which we used the fact that, at $\epsilon=0$, $g_1(x)|_{\epsilon=0}=1$, see\-~(\ref{g1const}).
In particular, the terms of order $0$ in $\epsilon$ are independent of $\xi$.
Note, in addition, that, by\-~(\ref{g11}),
\begin{equation}
  \int\frac{dx}V\ g_1^{(1)}(x)=0
  .
  \label{intg11}
\end{equation}
\bigskip

\point
The trick of this proof is to take the average with respect to $\xi$ on both sides of\-~(\ref{g2_xi}).
Since we take periodic boundary conditions, the $\Delta_\xi$ term drops out.
We will only focus on the first order contribution in $\epsilon$, and, as was mentioned above, terms of order $0$ are independent of $\xi$.
Thus, the average over $\xi$ will always apply to a single term, either $g_1^{(1)}$ or $u_2^{(1)}$.
By\-~(\ref{g11}), the terms involving $g_1^{(1)}$ have zero average.
We can therefore replace $g_1^{(1)}$ by 1.
(The previous argument does not apply to the terms in which $\Delta_\zeta$ acts on $g_1$, but these terms have a vanishing average as well because of the periodic boundary conditions.)
In particular, by\-~(\ref{g2g1}) and Lemma\-~\ref{lemma:g2},
\begin{equation}
  \int\frac{d\xi}V\ (1-u_2^{(1)}(\xi+{\textstyle\frac\zeta2},\xi-{\textstyle\frac\zeta 2}))
  =1
\end{equation}
so
\begin{equation}
  \int\frac{d\xi}V\ u_2^{(1)}(\xi+{\textstyle\frac\zeta2},\xi-{\textstyle\frac\zeta 2})
  =0
\end{equation}
and thus, we can replace $u_2$ with $u_2^{(0)}$.
Thus, using the translation invariant computation detailed in Section\-~\ref{sec:trsl_inv}, we find that the average of\-~(\ref{g2_xi}) is
\begin{equation}
  (-\Delta+v(\zeta)-2\rho K(\zeta)+\rho^2 L(\zeta))(1-u_2^{(0)}(\zeta))+\epsilon F(\zeta)+O(\epsilon^2)+\Sigma_2=0
  \label{eqNk_inproof}
\end{equation}
where $K$ and $L$ are defined in\-~(\ref{K}) and\-~(\ref{L}) and $F$ comes from the contribution to $\bar R_2$ of $\varpi$, see\-~(\ref{R}):
\begin{equation}
  \begin{largearray}
    F(\zeta):=\epsilon^{-1}\int \frac{d\xi}V\ 
    \left(
      \varpi_x+\varpi_y-2\left<\varpi\right>
      +\rho\int dz\ \varpi_z(u_2^{(0)}(\xi+{\textstyle\frac\zeta2}-z)u_2^{(0)}(\xi-{\textstyle\frac\zeta 2}-z))
      -\right.\\\hfill\left.-
      \rho\int dz\ \varpi_zu_2^{(0)}(\xi+{\textstyle\frac\zeta2}-z)
      -\rho\int dz\ \varpi_zu_2^{(0)}(\xi-{\textstyle\frac\zeta 2}-z)
    \right)(1-u_2^{(0)}(\zeta))
    .
  \end{largearray}
\end{equation}
Similarly, (\ref{Nken}) is
\begin{equation}
  e=\frac\rho2\int d\zeta\ v(\zeta)(1-u_2^{(0)}(\zeta))
  +\int\frac{dx}V\ \varpi g_1(x)
  +\Sigma_1
  +O(\epsilon^2)
  .
\end{equation}
\bigskip

\point
Furthermore, by\-~(\ref{varpiNk}),
\begin{equation}
  \int dz\ \varpi_z f(z)=0
\end{equation}
for any integrable $f$, so
\begin{equation}
  F(\zeta)=\epsilon^{-1}\int \frac{d\xi}V\ 
  \left(\varpi_x+\varpi_y\right)(1-u_2^{(0)}(\zeta))
\end{equation}
and
\begin{equation}
  e=\frac\rho2\int d\zeta\ v(\zeta)(1-u_2^{(0)}(\zeta))
  +\Sigma_1
  +O(\epsilon^2)
  .
  \label{Nken_inproof}
\end{equation}
Now,
\begin{equation}
  \varpi_x f(x-y)
  =
  e^{ikx}
  \int dz\ 
  e^{-ikz}f(z-y)
\end{equation}
so
\begin{equation}
  \varpi_x f(\zeta)
  =
  \epsilon e^{ik(\xi+{\textstyle\frac\zeta2})}
  \int dz\ 
  e^{-ik(z+(\xi-{\textstyle\frac\zeta 2}))}f(z)
  =
  \epsilon e^{ik\zeta}
  \int dz\ 
  e^{-ikz}f(z)
  =\epsilon e^{ik\zeta}\hat f(-k)
  .
\end{equation}
Similarly,
\begin{equation}
  \varpi_y f(\zeta)
  =\epsilon e^{-ik\zeta}\hat f(-k)
  .
\end{equation}
Thus
\begin{equation}
  F(\zeta)=2\cos(k\zeta)(\delta(k)-\hat u_2^{(0)}(-k))
  .
  \label{F_inproof}
\end{equation}
Since $k\neq 0$, the $\delta$ function drops out.
We conclude the proof by combining\-~(\ref{eqNk_inproof}), (\ref{Nken_inproof}) and\-~(\ref{F_inproof}) and taking the thermodynamic limit.
\qed

\subsection{The simple equation and Bogolyubov theory, proof of Theorem \expandonce{\ref{theo:Nk_bog}}}\label{sec:Nk_bog}

\point
We differentiate\-~(\ref{simpleq}) with respect to $\epsilon$ and take $\epsilon=0$:
\begin{equation}
  (-\Delta+v+4e+4e\rho u\ast)\partial_\epsilon u=-4\partial_\epsilon eu+2\partial_\epsilon e\rho u\ast u+F
  .
\end{equation}
Let
\begin{equation}
  \mathfrak K_e:=(-\Delta+v+4e(1-\rho u\ast))^{-1}
\end{equation}
(this operator was introduced and studied in detail in\-~\cite{CJL21}).
We apply $\mathfrak K_e$ to both sides and take a scalar product with $-\rho v/2$ and find
\begin{equation}
  \partial_\epsilon e=\rho\partial_\epsilon e\int dx\ v(x)\mathfrak K_e(2u(x)-\rho u\ast u(x))-\frac\rho2\int dx\ v(x)\mathfrak K_eF(x)
\end{equation}
and so, using\-~(\ref{M_simpleq}),
\begin{equation}
  \mathcal M^{(\mathrm{simpleq})}(k)=\partial_\epsilon e
  =-\frac{\frac\rho2\int dx\ v(x)\mathfrak K_eF(x)}{1-\rho\int dx\ v(x)\mathfrak K_e(2u(x)-\rho u\ast u(x))}
\end{equation}
and, by\-~(\ref{F}),
\begin{equation}
  \mathcal M^{(\mathrm{simpleq})}(k)
  =\rho\frac{\hat u(k)\int dx\ v(x)\mathfrak K_e\cos(kx)}{1-\rho\int dx\ v(x)\mathfrak K_e(2u(x)-\rho u\ast u(x))}
  .
\end{equation}
Note that
\begin{equation}
  \int\frac{dk}{(2\pi)^3}\mathcal M^{(\mathrm{simpleq})}(k)
  =
  \frac{\rho\int dx\ v(x)\mathfrak K_e u(x)}{1-\rho\int dx\ v(x)\mathfrak K_e(2u(x)-\rho u\ast u(x))}
\end{equation}
which is the expression for the uncondensed fraction for the simple equation\-~\cite[(38)]{CHe21}.
\bigskip

\point
By\-~\cite[(5.8),(5.27)]{CJL21},
\begin{equation}
  \mathcal M^{(\mathrm{simpleq})}(k)=\rho
  \left(\hat u(k)\int dx\ v(x)\mathfrak K_e\cos(k(x))\right)
  (1+O(\rho e^{-\frac12}))
  .
\end{equation}
Furthermore, by the resolvent identity,
\begin{equation}
  \mathfrak K_e\cos(kx)
  =
  \xi-\mathfrak K_e(v\xi)
  ,\quad
  \xi:=\mathfrak Y_e(\cos(kx))
  :=(-\Delta+4e(1-\rho u\ast))^{-1}\cos(kx)
\end{equation}
in terms of which, using the self-adjointness of $\mathfrak K_e$,
\begin{equation}
  \mathcal M^{(\mathrm{simpleq})}(k)=\rho\hat u(k)\left(
    \int dx\ v(x)\xi(x)
    -
    \int dx\ \mathfrak K_ev(x)(v(x)\xi(x))
  \right)
  .
  \label{pde}
\end{equation}
\bigskip

\point
Now, taking the Fourier transform,
\begin{equation}
  \hat\xi(q)\equiv\int dx\ e^{ikx}\xi(x)=\frac{(2\pi)^3}2\frac{\delta(k-q)+\delta(k+q)}{q^2+4e(1-\rho\hat u(q))}
\end{equation}
and so
\begin{equation}
  \int dx\ v(x)\xi(x)
  =
  \int\frac{dq}{(2\pi)^3}\hat v(q)\hat\xi(q)
  =
  \frac{\hat v(k)}{k^2+4e(1-\rho\hat u(k))}
\end{equation}
and thus
\begin{equation}
  \rho\hat u(k)\int dx\ v(x)\xi
  =
  \rho\hat v(k)\frac{\hat u(k)}{k^2+4e(1-\rho\hat u(k))}
  .
\end{equation}
We recall\-~\cite[(4.25)]{CJL20}:
\begin{equation}
  \rho\hat u(k)=\frac{k^2}{4e}+1-\sqrt{\left(\frac{k^2}{4e}+1\right)^2-\hat S(k)}
  \label{rhou}
\end{equation}
and, by\-~\cite[(4.24)]{CJL20},
\begin{equation}
  \hat S(0)=1
  .
  \label{S1}
\end{equation}
Therefore, if we rescale
\begin{equation}
  k=2\sqrt{e}\kappa
\end{equation}
we find
\begin{equation}
  \rho\hat u(k)\int dx\ v(x)\xi
  =
  \frac{\hat v(0)}{4e}\frac{\kappa^2+1-\sqrt{(\kappa^2+1)^2-1}}{\sqrt{(\kappa^2+1)^2-1}}
  +o(e^{-1})
  .
  \label{pde1}
\end{equation}
\bigskip

\point
Now,
\begin{equation}
  \int dx\ e^{iqx}v(x)\xi(x)
  =
  \frac12\frac1{k^2+4e(1-\rho\hat u(k))}
  \int dp\ \hat v(q-p)(\delta(k-p)+\delta(k+p))
\end{equation}
so
\begin{equation}
  \int dx\ e^{iqx}v(x)\xi(x)
  =
  \frac12\frac{\hat v(q-k)+\hat v(q+k)}{k^2+4e(1-\rho\hat u(k))}
  .
\end{equation}
Therefore,
\begin{equation}
  \int dx\ \mathfrak K_ev(x)(v\xi)
  =
  \frac12\frac1{k^2+4e(1-\rho\hat u(k))}
  \int\frac{dq}{(2\pi)^3}\ 
  \widehat{\mathfrak K_e v}(q)
  (\hat v(k-q)+\hat v(k+q))
\end{equation}
which, using the $q\mapsto-q$ symmetry, is
\begin{equation}
  \int dx\ \mathfrak K_ev(x)(v\xi)
  =
  \frac1{k^2+4e(1-\rho\hat u(k))}
  \int\frac{dq}{(2\pi)^3}\ 
  \widehat{\mathfrak K_e v}(q)
  \hat v(k+q)
\end{equation}
that is,
\begin{equation}
  \rho\hat u(k)\int dx\ \mathfrak K_ev(x)(v\xi)
  =
  \frac{\rho\hat u(k)}{k^2+4e(1-\rho\hat u(k))}
  \int dx\ 
  e^{-ikx}
  \mathfrak K_e v(x)
  v(x)
\end{equation}
in which we rescale
\begin{equation}
  k=2\sqrt e\kappa
\end{equation}
so, by\-~(\ref{rhou})-(\ref{S1}),
\begin{equation}
  \rho\hat u(k)\int dx\ \mathfrak K_ev(x)(v\xi)
  =
  \frac{\kappa^2+1-\sqrt{(\kappa^2+1)^2-1}}{4e\sqrt{(\kappa^2+1)^2-1}}
  (1+o(1))\int dx\ 
  e^{-i2\sqrt e\kappa x}
  v(x)\mathfrak K_e v(x)
  .
\end{equation}
Therefore, by dominated convergence (using the argument above\-~\cite[(5.23)]{CJL21} and the fact that $\mathfrak K_e$ is positivity preserving), and by\-~\cite[(5.23)-(5.24)]{CJL21},
\begin{equation}
  \rho\hat u(k)\int dx\ \mathfrak K_ev(x)(v\xi)
  =
  \frac{\kappa^2+1-\sqrt{(\kappa^2+1)^2-1}}{4e\sqrt{(\kappa^2+1)^2-1}}
  (-4\pi a+\hat v(0))+o(e^{-1})
  .
  \label{pde2}
\end{equation}
\bigskip

\point
Inserting\-~(\ref{pde1}) and\-~(\ref{pde2}) into\-~(\ref{pde}), we find
\begin{equation}
  \mathcal M^{(\mathrm{simpleq})}(k)
  =
  \frac{\pi a}{e}\frac{\kappa^2+1-\sqrt{(\kappa^2+1)^2-1}}{\sqrt{(\kappa^2+1)^2-1}}
  +o(e^{-1})
  .
\end{equation}
Finally, we recall\-~\cite[(1.23)]{CJL20}:
\begin{equation}
  e=2\pi\rho a(1+O(\sqrt\rho))
  \label{erho}
\end{equation}
so
\begin{equation}
  \mathcal M^{(\mathrm{simpleq})}(k)
  =
  \frac{1}{2}\frac{\kappa^2+1-\sqrt{(\kappa^2+1)^2-1}}{\sqrt{(\kappa^2+1)^2-1}}
  +o(e^{-1})
  .
  \label{final1}
\end{equation}
\bigskip

\point
Finally, by\-~(\ref{Mbog})
\begin{equation}
  \mathcal M^{(\mathrm{Bogolyubov})}(2\sqrt e\kappa)=-\frac1{2\rho}\left(1-\frac{\frac{4e}{8\pi\rho a}\kappa^2+1}{\sqrt{\frac{e^2}{4\pi^2\rho^2a^2}\kappa^4+\frac{e}{\pi\rho a} \kappa^2}}\right)
\end{equation}
so by\-~(\ref{erho}),
\begin{equation}
  \mathcal M^{(\mathrm{Bogolyubov})}(2\sqrt e\kappa)=-\frac1{2\rho}\left(1-\frac{\kappa^2+1}{\sqrt{\kappa^4+2\kappa^2}}\right)
  .
\end{equation}
This, together with\-~(\ref{final1}), implies\-~(\ref{Msimpleqbog}).
\qed

\vfill
\eject


\begin{thebibliography}{WWW99}
\small
\IfFileExists{bibliography/bibliography.tex}{\bibitem[BCS21]{BCS21}G. Basti, S. Cenatiempo, B. Schlein - {\it A new second-order upper bound for the ground state energy of dilute Bose gases}, Forum of Mathematics, Sigma, volume\-~9, number e74, 2021,\par\penalty10000
doi:{\tt\color{blue}\href{http://dx.doi.org/10.1017/fms.2021.66}{10.1017/fms.2021.66}}, arxiv:{\tt\color{blue}\href{http://arxiv.org/abs/2101.06222}{2101.06222}}.\par\medskip
 
\bibitem[BBe18]{BBe18}C. Boccato, C. Brennecke, S. Cenatiempo, B. Schlein - {\it Complete Bose–Einstein Condensation in the Gross–Pitaevskii Regime}, Communications in Mathematical Physics, volume\-~359, issue\-~3, pages\-~975-1026, 2018,\par\penalty10000
doi:{\tt\color{blue}\href{http://dx.doi.org/10.1007/s00220-017-3016-5}{10.1007/s00220-017-3016-5}}.\par\medskip
 
\bibitem[BBe19]{BBe19}C. Boccato, C. Brennecke, S. Cenatiempo, B. Schlein - {\it Bogoliubov theory in the Gross–Pitaevskii limit}, Acta Mathematica, volume\-~222, issue\-~2, pages\-~219-335, 2019,\par\penalty10000
doi:{\tt\color{blue}\href{http://dx.doi.org/10.4310/ACTA.2019.v222.n2.a1}{10.4310/ACTA.2019.v222.n2.a1}}, arxiv:{\tt\color{blue}\href{http://arxiv.org/abs/1801.01389}{1801.01389}}.\par\medskip
 
\bibitem[BBe20]{BBe20}C. Boccato, C. Brennecke, S. Cenatiempo, B. Schlein - {\it Optimal Rate for Bose-Einstein Condensation in the Gross-Pitaevskii Regime}, Communications in Mathematical Physics, volume\-~376, issue\-~2, pages\-~1311-1395, 2020,\par\penalty10000
doi:{\tt\color{blue}\href{http://dx.doi.org/10.1007/s00220-019-03555-9}{10.1007/s00220-019-03555-9}}, arxiv:{\tt\color{blue}\href{http://arxiv.org/abs/1812.03086}{1812.03086}}.\par\medskip
 
\bibitem[Bo47]{Bo47}N. Bogolubov - {\it On the theory of superfluidity}, Journal of Physics (USSR), volume\-~11, number\-~1, pages\-~23-32 (translated from the Russian Izv.Akad.Nauk Ser.Fiz, volume\-~11, pages\-~77-90), 1947.\par\medskip
 
\bibitem[BSS22]{BSS22}C. Brennecke, B. Schlein, S. Schraven - {\it Bose-Einstein Condensation with Optimal Rate for Trapped Bosons in the Gross-Pitaevskii Regime}, Mathematical Physics, Analysis and Geometry, volume\-~25, issue\-~2, pages\-~1-71, 2022,\par\penalty10000
doi:{\tt\color{blue}\href{http://dx.doi.org/10.1007/s11040-022-09424-7}{10.1007/s11040-022-09424-7}}, arxiv:{\tt\color{blue}\href{http://arxiv.org/abs/2102.11052}{2102.11052}}.\par\medskip
 
\bibitem[BSS22b]{BSS22b}C. Brennecke, B. Schlein, S. Schraven - {\it Bogoliubov Theory for Trapped Bosons in the Gross-Pitaevskii Regime}, Annales Henri Poincaré, volume\-~23, issue\-~5, pages\-~1583-1658, 2022,\par\penalty10000
doi:{\tt\color{blue}\href{http://dx.doi.org/10.1007/s00023-021-01151-z}{10.1007/s00023-021-01151-z}}, arxiv:{\tt\color{blue}\href{http://arxiv.org/abs/2108.11129}{2108.11129}}.\par\medskip
 
\bibitem[CHe21]{CHe21}E.A. Carlen, M. Holzmann, I. Jauslin, E.H. Lieb - {\it Simplified approach to the repulsive Bose gas from low to high densities and its numerical accuracy}, Physical Review A, volume\-~103, issue\-~5, number\-~053309, 2021,\par\penalty10000
doi:{\tt\color{blue}\href{http://dx.doi.org/10.1103/PhysRevA.103.053309}{10.1103/PhysRevA.103.053309}}, arxiv:{\tt\color{blue}\href{http://arxiv.org/abs/2011.10869}{2011.10869}}.\par\medskip
 
\bibitem[CJL20]{CJL20}E.A. Carlen, I. Jauslin, E.H. Lieb - {\it Analysis of a simple equation for the ground state energy of the Bose gas}, Pure and Applied Analysis, volume\-~2, issue\-~3, pages\-~659-684, 2020,\par\penalty10000
doi:{\tt\color{blue}\href{http://dx.doi.org/10.2140/paa.2020.2.659}{10.2140/paa.2020.2.659}}, arxiv:{\tt\color{blue}\href{http://arxiv.org/abs/1912.04987}{1912.04987}}.\par\medskip
 
\bibitem[CJL21]{CJL21}E.A. Carlen, I. Jauslin, E.H. Lieb - {\it Analysis of a Simple Equation for the Ground State of the Bose Gas II: Monotonicity, Convexity, and Condensate Fraction}, SIAM Journal on Mathematical Analysis, volume\-~53, number\-~5, pages\-~5322-5360, 2021,\par\penalty10000
doi:{\tt\color{blue}\href{http://dx.doi.org/10.1137/20M1376820}{10.1137/20M1376820}}, arxiv:{\tt\color{blue}\href{http://arxiv.org/abs/2010.13882}{2010.13882}}.\par\medskip
 
\bibitem[DS20]{DS20}A. Deuchert, R. Seiringer - {\it Gross-Pitaevskii Limit of a Homogeneous Bose Gas at Positive Temperature}, Archive for Rational Mechanics and Analysis, volume\-~236, issue\-~3, pages\-~1217-1271, 2020,\par\penalty10000
doi:{\tt\color{blue}\href{http://dx.doi.org/10.1007/s00205-020-01489-4}{10.1007/s00205-020-01489-4}}, arxiv:{\tt\color{blue}\href{http://arxiv.org/abs/1901.11363}{1901.11363}}.\par\medskip
 
\bibitem[DSY19]{DSY19}A. Deuchert, R. Seiringer, J. Yngvason - {\it Bose-Einstein Condensation in a Dilute, Trapped Gas at Positive Temperature}, Communications in Mathematical Physics, volume\-~368, issue\-~2, pages\-~723-776, 2019,\par\penalty10000
doi:{\tt\color{blue}\href{http://dx.doi.org/10.1007/s00220-018-3239-0}{10.1007/s00220-018-3239-0}}, arxiv:{\tt\color{blue}\href{http://arxiv.org/abs/1803.05180}{1803.05180}}.\par\medskip
 
\bibitem[Dy57]{Dy57}F.J. Dyson - {\it Ground-State Energy of a Hard-Sphere Gas}, Physical Review, volume\-~106, issue\-~1, pages\-~20-26, 1957,\par\penalty10000
doi:{\tt\color{blue}\href{http://dx.doi.org/10.1103/PhysRev.106.20}{10.1103/PhysRev.106.20}}.\par\medskip
 
\bibitem[FS20]{FS20}S. Fournais, J.P. Solovej - {\it  The energy of dilute Bose gases}, Annals of Mathematics, volume\-~192, issue\-~3, pages\-~893-976, 2020,\par\penalty10000
doi:{\tt\color{blue}\href{http://dx.doi.org/10.4007/annals.2020.192.3.5}{10.4007/annals.2020.192.3.5}}, arxiv:{\tt\color{blue}\href{http://arxiv.org/abs/1904.06164}{1904.06164}}.\par\medskip
 
\bibitem[FS22]{FS22}S. Fournais, J.P. Solovej - {\it The energy of dilute Bose gases II: the general case}, Inventiones mathematicae, volume , issue , pages\-~1-132, 2022,\par\penalty10000
doi:{\tt\color{blue}\href{http://dx.doi.org/10.1007/s00222-022-01175-0}{10.1007/s00222-022-01175-0}}, arxiv:{\tt\color{blue}\href{http://arxiv.org/abs/2108.12022}{2108.12022}}.\par\medskip
 
\bibitem[Gr61]{Gr61}E.P. Gross - {\it Structure of a quantized vortex in boson systems}, Il Nuovo Cimento (1955-1965), volume\-~20, issue\-~3, pages\-~454-477, 1961,\par\penalty10000
doi:{\tt\color{blue}\href{http://dx.doi.org/10.1007/BF02731494}{10.1007/BF02731494}}.\par\medskip
 
\bibitem[HST22]{HST22}C.\-~Hainzl, B.\-~Schlein, A.\-~Triay - {\it Bogoliubov theory in the Gross-Pitaevskii limit}, arXiv preprint, 2022\par\penalty10000
arxiv:{\tt\color{blue}\href{https://arxiv.org/abs/2203.03440}{2203.03440}}.\par\medskip
 
\bibitem[Ja22]{Ja22}I. Jauslin - {\it Review of a Simplified Approach to study the Bose gas at all densities}, The Physics and Mathematics of Elliott Lieb, The\-~90th Anniversary Volume I, chapter\-~25, pages\-~609-635, ed. Rupert L. Frank, Ari Laptev, Mathieu Lewin, Robert Seiringer, EMS Press, 2022,\par\penalty10000
doi:{\tt\color{blue}\href{http://dx.doi.org/10.4171/90-1/25}{10.4171/90-1/25}}, arxiv:{\tt\color{blue}\href{http://arxiv.org/abs/2202.07637}{2202.07637}}.\par\medskip
 
\bibitem[LHY57]{LHY57}T.D. Lee, K. Huang, C.N. Yang - {\it Eigenvalues and Eigenfunctions of a Bose System of Hard Spheres and Its Low-Temperature Properties}, Physical Review, volume\-~106, issue\-~6, pages\-~1135-1145, 1957,\par\penalty10000
doi:{\tt\color{blue}\href{http://dx.doi.org/10.1103/PhysRev.106.1135}{10.1103/PhysRev.106.1135}}.\par\medskip
 
\bibitem[Li63]{Li63}E.H. Lieb - {\it Simplified Approach to the Ground-State Energy of an Imperfect Bose Gas}, Physical Review, volume\-~130, issue\-~6, pages\-~2518-2528, 1963,\par\penalty10000
doi:{\tt\color{blue}\href{http://dx.doi.org/10.1103/PhysRev.130.2518}{10.1103/PhysRev.130.2518}}.\par\medskip
 
\bibitem[LL64]{LL64}E.H. Lieb, W. Liniger - {\it Simplified Approach to the Ground-State Energy of an Imperfect Bose Gas. III. Application to the One-Dimensional Model}, Physical Review, volume\-~134, issue\-~2A, pages A312-A315, 1964,\par\penalty10000
doi:{\tt\color{blue}\href{http://dx.doi.org/10.1103/PhysRev.134.A312}{10.1103/PhysRev.134.A312}}.\par\medskip
 
\bibitem[LS64]{LS64}E.H. Lieb, A.Y. Sakakura - {\it Simplified Approach to the Ground-State Energy of an Imperfect Bose Gas. II. Charged Bose Gas at High Density}, Physical Review, volume\-~133, issue\-~4A, pages A899-A906, 1964,\par\penalty10000
doi:{\tt\color{blue}\href{http://dx.doi.org/10.1103/PhysRev.133.A899}{10.1103/PhysRev.133.A899}}.\par\medskip
 
\bibitem[LS02]{LS02}E.H. Lieb, R. Seiringer - {\it Proof of Bose-Einstein Condensation for Dilute Trapped Gases}, Physical Review Letters, volume\-~88, issue\-~17, number\-~170409, 2002,\par\penalty10000
doi:{\tt\color{blue}\href{http://dx.doi.org/10.1103/PhysRevLett.88.170409}{10.1103/PhysRevLett.88.170409}}, arxiv:{\tt\color{blue}\href{http://arxiv.org/abs/math-ph/0112032}{math-ph/0112032}}.\par\medskip
 
\bibitem[LSe05]{LSe05}E.H. Lieb, R. Seiringer, J.P. Solovej, J. Yngvason - {\it The Mathematics of the Bose Gas and its Condensation}, Oberwolfach Seminars, volume\-~34, Birkha\"user, 2005, arxiv:{\tt\color{blue}\href{http://arxiv.org/abs/cond-mat/0610117}{cond-mat/0610117}}.\par\medskip
 
\bibitem[LSY00]{LSY00}E.H. Lieb, R. Seiringer, J. Yngvason - {\it Bosons in a trap: A rigorous derivation of the Gross-Pitaevskii energy functional}, Physical Review A, volume\-~61, issue\-~4, number\-~043602, 2000,\par\penalty10000
doi:{\tt\color{blue}\href{http://dx.doi.org/10.1103/PhysRevA.61.043602}{10.1103/PhysRevA.61.043602}}, arxiv:{\tt\color{blue}\href{http://arxiv.org/abs/math-ph/9908027}{math-ph/9908027}}.\par\medskip
 
\bibitem[LY98]{LY98}E.H. Lieb, J. Yngvason - {\it Ground State Energy of the Low Density Bose Gas}, Physical Review Letters, volume\-~80, issue\-~12, pages\-~2504-2507, 1998,\par\penalty10000
doi:{\tt\color{blue}\href{http://dx.doi.org/10.1103/PhysRevLett.80.2504}{10.1103/PhysRevLett.80.2504}}, arxiv:{\tt\color{blue}\href{http://arxiv.org/abs/cond-mat/9712138}{cond-mat/9712138}}.\par\medskip
 
\bibitem[NNe22]{NNe22}P.T. Nam, M. Napi\'orkowski, J. Ricaud, A. Triay - {\it Optimal rate of condensation for trapped bosons in the Gross–Pitaevskii regime}, Analysis and PDE, volume\-~15, issue\-~6, pages\-~1585-1616, 2022,\par\penalty10000
doi:{\tt\color{blue}\href{http://dx.doi.org/10.2140/apde.2022.15.1585}{10.2140/apde.2022.15.1585}}, arxiv:{\tt\color{blue}\href{http://arxiv.org/abs/2001.04364}{2001.04364}}.\par\medskip
 
\bibitem[NRS16]{NRS16}P.T. Nam, N. Rougerie, R. Seiringer - {\it Ground states of large bosonic systems : the Gross–Pitaevskii limit revisited}, Analysis and PDE, volume\-~9, issue\-~2, pages\-~459-485, 2016,\par\penalty10000
doi:{\tt\color{blue}\href{http://dx.doi.org/10.2140/apde.2016.9.459}{10.2140/apde.2016.9.459}}, arxiv:{\tt\color{blue}\href{http://arxiv.org/abs/1503.07061}{1503.07061}}.\par\medskip
 
\bibitem[NT21]{NT21}P.T.\-~Nam, A.\-~Triay - {\it Bogoliubov excitation spectrum of trapped Bose gases in the Gross-Pitaevskii regime}, arXiv preprint, 2021\par\penalty10000
arxiv:{\tt\color{blue}\href{https://arxiv.org/abs/2106.11949}{2106.11949}}.\par\medskip
 
\bibitem[Pi61]{Pi61}L.P. Pitaevskii - {\it Vortex lines in an imperfect Bose gas}, Soviet Physics JETP, volume\-~13, number\-~2, pages\-~451-454, 1961.\par\medskip
 
\bibitem[Sc22]{Sc22}B. Schlein - {\it Bose gases in the Gross-Pitaevskii limit: A survey of some rigorous results}, The Physics and Mathematics of Elliott Lieb, The\-~90th anniversary volume II, eds R.L. Frank, A. Laptev, M. Lewin, R. Seiringer, pages\-~277-305, 2022,\par\penalty10000
doi:{\tt\color{blue}\href{http://dx.doi.org/10.4171/90-2/40}{10.4171/90-2/40}}, arxiv:{\tt\color{blue}\href{http://arxiv.org/abs/2203.10855}{2203.10855}}.\par\medskip
 
\bibitem[YY09]{YY09}H. Yau, J. Yin - {\it The Second Order Upper Bound for the Ground Energy of a Bose Gas}, Journal of Statistical Physics, volume\-~136, issue\-~3, pages\-~453-503, 2009,\par\penalty10000
doi:{\tt\color{blue}\href{http://dx.doi.org/10.1007/s10955-009-9792-3}{10.1007/s10955-009-9792-3}}, arxiv:{\tt\color{blue}\href{http://arxiv.org/abs/0903.5347}{0903.5347}}.\par\medskip
 
}{}
\end{thebibliography}
\end{document}